\documentclass[12pt,preprint]{aastex}

\newcommand{\km}{\,{\textnormal{km}}}
\newcommand{\second}{\,{\textnormal{s}}}
\newcommand{\kms}{{\km\second^{-1}}}
\newcommand{\solarmass}{{\,{\textnormal{M}}_\sun}}
\newcommand{\yr}{{\,\textnormal{yr}}}
\newcommand{\solarmassyear}{{\solarmass\yr^{-1}}}
\newcommand{\AU}{{\,\textnormal{AU}}}
\newcommand{\stellarmass}{{{\textnormal{M}}_\ast}}

\newcommand{\kmsns}{{{\textnormal{km}}\second^{-1}}}
\newcommand{\AUns}{{\textnormal{AU}}}
\newcommand{\secondns}{{\textnormal{s}}}

\begin{document}

\shorttitle{LAUNCHING REGION OF T TAURI WINDS}
\shortauthors{ANDERSON ET AL.}

\title{Locating the Launching Region of T Tauri Winds:\\  The Case of DG Tau}

\author{Jeffrey M. Anderson and Zhi-Yun Li}
\affil{Department of Astronomy, University of Virginia, P. O. Box 3818,
 Charlottesville, VA 22903}
\email{jma2u@virginia.edu, zl4h@virginia.edu}
\author{Ruben Krasnopolsky}
\affil{Center for Theoretical Astrophysics, University of Illinois at
Urbana-Champaign, Loomis Laboratory, 1110 West Green Street,
Urbana, IL 61801}
\email{ruben@astro.uiuc.edu}
\author{Roger D. Blandford}
\affil{California Institute of Technology, Theoretical Astrophysics,
130-30 Caltech, Pasadena, CA 91125}
\email{rdb@tapir.caltech.edu}

\begin{abstract}
It is widely believed that T Tauri winds are driven magnetocentrifugally
from accretion disks close to the central stars. The exact launching
conditions are uncertain. We show that a general relation exists between
the poloidal and toroidal velocity components of a magnetocentrifugal
wind at large distances and the rotation rate of the launching surface,
independent of the uncertain launching conditions. We discuss the
physical basis of this relation and verify it using a set of
numerically-determined large-scale wind solutions. Both velocity
components are in principle measurable from spatially resolved spectra,
as has been done for the extended low-velocity component (LVC) of the
DG Tau wind by Bacciotti et al. For this particular source, we infer
that the spatially resolved LVC originates from a region on the disk
extending from $\sim 0.3$ to $\sim 4.0\AU$ from the star, which is
consistent with, and a refinement over, the previous rough estimate of
Bacciotti et al.
\end{abstract}


\section{INTRODUCTION}
\label{sec:introduction}

The magnetocentrifugal mechanism \citep{bp82} is widely considered the
leading candidate for producing jets and winds in star formation (see
reviews by \citealt{kp00}; \citealt{shu00}); alternative mechanisms involving
thermal and/or radiation pressure have been shown to be inadequate. An
outflow can in principle be launched by torsional Alfv{\'e}n waves, as
shown numerically by \citet{su85} and others (see \citealt{kks02} and
references therein). These simulations
are typically of relatively short duration, however, and it is not
clear whether a large-scale, sustained outflow can be produced by this
mechanism alone.

The magnetocentrifugal model has two popular versions: X-winds \citep{shu00}
and disk-winds \citep{kp00}. The X-winds are launched from a narrow
region close to the truncation radius $R_X$ of the Keplerian disk by a
stellar magnetosphere. The disk-winds are, on the other hand, envisioned
to come from a wider range of disk radii outside $R_X$. These two
possibilities are not immediately distinguishable observationally, because
the launching surfaces in both cases are thought to be small, and only
the flow regions at much larger distances are directly observable at the
present.

In this Letter, we derive a method of inferring the rotation rate (and
thus the radius) of the wind-launching region on the Keplerian disk
from measurable quantities at large distances (\S\,\ref{sec:method}).
In \S\,\ref{sec:application}, we
first test the method on a set of numerically-determined large-scale
magnetocentrifugal wind solutions, and then apply it to the well-studied
wind of the T Tauri star DG Tau, using data obtained by \citet{bac02}
with the Space Telescope Imaging Spectrograph (STIS) of the
\textit{Hubble Space Telescope} (HST). We discuss our results and
conclude in \S\,\ref{sec:conclusions}.

\section{A METHOD FOR LOCATING THE WIND-LAUNCHING REGION}
\label{sec:method}

\subsection{Physical Basis of the Method}

The basic principle of magnetocentrifugal wind launching is well understood.
It involves open field lines firmly anchored
on a rapidly rotating disk, and centrifugal acceleration of fluid parcels
along the field lines into a high-speed wind. The wind trails behind the
disk in rotation, generating a toroidal component of magnetic field, which
exerts a braking torque on the disk. It is this magnetic torque that is
responsible for extracting both energy and angular momentum from the disk
and for powering the wind. Since the energy extracted is simply the work
done by the rotating disk against the magnetic torque (e.g.\ \citealt{s96}),
the rate of energy extraction is directly proportional to the rate of
angular momentum extraction, with the proportionality constant being the
angular speed $\Omega_0$ of the disk rotation. The extracted energy and
angular momentum are initially stored in an electromagnetic form. They
are gradually converted into a kinetic form as the flow accelerates. At
large observable distances, the conversion is nearly complete, and the
wind becomes kinetically dominated. Along any given field line, the
kinetic energy will then be proportional to the fluid angular momentum,
with the same proportionality constant $\Omega_0$ because of energy and
angular momentum conservation.

At any observable location (say of a distance $\varpi_\infty$ from the
rotation axis, where the subscript denotes a quantity far from the
launching region), the specific kinetic energy and angular momentum
of the wind are given by the poloidal and toroidal components of the
fluid velocity, $v_{p,\infty}$ and $v_{\phi,\infty}$, through $(v_{p,
\infty}^2+v_{\phi,\infty}^2)/2$ and $v_{\phi,\infty}\varpi_\infty$
respectively. Both velocity components can in principle be derived
empirically from spatially resolved spectra and proper motion observations,
as has been done for DG Tau \citep{bac02}. Once measured, they
can be used to deduce the rate of disk rotation at the foot point of the
field line passing through that location, through
\begin{equation}
\Omega_0\approx {v_{p,\infty}^2/2\over v_{\phi,\infty}\varpi_\infty},
\label{eqn:relation1}
\end{equation}
where $v_{\phi,\infty}\ll v_{p,\infty}$ is assumed, as is generally true
at large distances.
The above relation provides a simple way to infer the rotation rate
of the wind-launching region on the Keplerian disk and, if the central
stellar mass is known independently, the distance from the star. In the
next subsection, we will use the steady magnetohydrodynamic (MHD) wind
theory to derive a refined version of this relation, taking into account
of the energy and angular momentum associated with fluid rotation at the
base of the wind.

\subsection{General Derivation}

It is well known that in a steady, axisymmetric MHD wind, several quantities
are conserved along a given field line \citep{m68}. These include the
total specific energy and angular momentum
\begin{equation}
E = \frac{v^{2}}{2} - \frac{B_{\phi}B_{p}\Omega\varpi}{4\pi\rho v_{p}}
+ h + \Phi_{g}\,, \ \
L= \varpi\left(v_{\phi} - \frac{B_{\phi}B_{p}}{4\pi\rho v_{p}}\right),
\label{eqn:el}
\end{equation}
and the quantity
$
\Omega=(v_{\phi} - B_{\phi} v_{p}/B_{p})/\varpi
$,
which can be interpreted as the angular speed at the base of the wind,
$\Omega_0$, where the poloidal flow speed $v_p$ is negligible compared
to the speed of Keplerian rotation. Here, $v$ denotes velocity, $B$ magnetic
field, $h$ specific enthalpy, $\rho$ mass density, and $\Phi_{g}$ the
gravitational potential. The subscript $p$ refers to a quantity in the
poloidal $(\varpi, z)$ plane of a cylindrical coordinate system $(\varpi,
\phi,z)$. We will ignore the enthalpy term in the expression for
specific energy since magnetocentrifugal winds are dynamically cold in
general.

Note from equation (\ref{eqn:el}) that the magnetic contributions to the
energy and angular momentum are proportional to each other. Since neither
of them can be measured directly, we get rid of both by constructing a
combined quantity
\begin{equation}
J\equiv E-\Omega L = \frac{v^{2}}{2} + \Phi_{g} - \Omega\varpi v_{\phi},
\label{eqn:combined}
\end{equation}
which is also conserved along a field line (see also \citealt{lov86}).
At the launching surface, the wind corotates with the disk at the local
Keplerian speed $v_{K,0}$, which yields $J=-3 v_{K,0}^2/2$. As the
distance increases, the gravitational potential $\Phi_{g}$ decreases quickly,
and we have approximately
\begin{equation}
\frac{v_{p,\infty}^{2}+v_{\phi,\infty}^{2}}{2}
-\Omega_0 \varpi_{\infty}v_{\phi,\infty}\approx
-\frac{3}{2}v_{K,0}^{2}.
\label{eqn:final1}
\end{equation}
Since the Keplerian speed of disk rotation is related to the angular
speed through $v_{K,0}=(G\stellarmass\Omega_{0})^{1/3}$ around a star of mass
$\stellarmass$, we finally have
\begin{equation}
\varpi_{\infty}v_{\phi,\infty}\Omega_{0} - \frac{3}{2}(G\stellarmass)^{2/3}
\Omega_{0}^{2/3} - \frac{v_{p,\infty}^{2}+v_{\phi,\infty}^{2}}{2}
\approx 0,
\label{eqn:final2}
\end{equation}
which is the desired equation for determining the disk rotation rate
$\Omega_0$ in the wind-launching region from measurable quantities
at large distances.

Mathematically, we can define $\xi\equiv \Omega_{0}^{1/3}$, and cast
equation (\ref{eqn:final2}) into a cubic equation
\begin{equation}
\xi^{3}-a_{2}\xi^{2}-a_{0}=0,
\label{eqn:cubic}
\end{equation}
with the coefficients
\begin{equation}
a_{2}=\frac{3}{2}\frac{(G\stellarmass)^{2/3}}{\varpi_{\infty}v_{\phi,\infty}},
\ \ \ {\rm and}\ \ \
a_{0}=\frac{v_{p,\infty}^{2}+v_{\phi,\infty}^{2}}{2\varpi_{\infty}
v_{\phi,\infty}}.
\end{equation}
If we let $q=-a_{2}^{2}/9$, $r=a_{0}/2+a_{2}^{3}/27$, and $D=q^{3}+r^{2}
=a_{0}^{2}/4+a_{2}^{3}a_{0}/27$, then the nature of the solution is
determined by the value of $D$.  For $D>0$, which is always true in
our case, there are one real and two complex conjugate roots.
The real root is given analytically by
\begin{equation}
\xi=(r+\sqrt{D})^{1/3}+(r-\sqrt{D})^{1/3}+a_{2}/3.
\label{eqn:cubsoln}
\end{equation}
Once $\xi$ is determined, one can obtain the angular speed through
$\Omega_0=\xi^3$, and infer the wind-launching radius through
$\varpi_0= (G\stellarmass/\Omega_0^2)^{1/3}$.

Astrophysically interesting magnetocentrifugal winds are probably ``fast''
in the sense that they have enough energy to climb out the potential well
of the central star easily (instead of barely). In such a case, we expect
the kinetic energy of the wind (the third term of eq.\ [\ref{eqn:final2}])
to be substantially greater than the gravitational binding energy at the
launching surface (which is essentially the second term of
eq.\ [\ref{eqn:final2}]).
Noting again that typically $v_{\phi,\infty}\ll v_{p,
\infty}$, we recover from equation (\ref{eqn:final2}) the simple relation
(\ref{eqn:relation1}), which can be cast into a more practical form
\begin{equation}
\varpi_{0}\approx0.7\AU\left(\frac{\varpi_{\infty}}
{10\AU}\right)^{2/3}\left(
\frac{v_{\phi,\infty}}{10\kms}\right)^{2/3}
\left(\frac{v_{p,\infty}}{100\kms}\right)^{-4/3}
\left(\frac{\stellarmass}{1\solarmass}\right)^{1/3}
\label{eqn:final4}
\end{equation}
for locating the wind-launching region on the disk.

\section{METHOD VERIFICATION AND APPLICATION}
\label{sec:application}

\subsection{Method Verification using Numerical Solutions}

We have tested the method of inferring the rotation rate of the
wind-launching surface on a set of numerically-determined
magnetocentrifugal wind solutions (J.~Anderson et al.\ 2003, in
preparation). These solutions are obtained
using the {\sc{Zeus3D}} MHD code \citep{cnf94} assuming
axisymmetry. The simulation setup and code modifications are described
in \citet{klb99}. Basically, we prescribe on the Keplerian disk around
a solar-mass star a distribution of open magnetic field, and load onto
the field a mass flux at a low speed (typically $10\%$ of Keplerian).
The slowly-moving wind material is accelerated centrifugally along the
rotating field lines into a high-speed flow. The wind gradually sweeps
the ambient medium out of the simulation box, chosen to have a $100\times
100\AU$ size, and settles into a steady state. The properties of the
steady wind at large distances are determined uniquely by the conditions
on the launching surface. They are used for our method verification.

The results of applying our method to two representative simulations
are presented in Table~\ref{tab:model}.
The simulations differ only in the mass loading rate of the wind; model
L corresponds to a ``light'' wind, with $\dot{M}_{w} = 10^{-8}\solarmassyear$,
typical of T Tauri winds, and model H corresponds
to a ``heavy'' wind, with $\dot{M}_{w} = 10^{-6}\solarmassyear$,
which is more extreme \citep{erm93}. For each model, we first
read out from the solution the two velocity components $v_{p,\infty}$
and $v_{\phi,\infty}$ at four selected, observable locations with
known cylindrical radius $\varpi_\infty$, and then compute the
corresponding angular speed at the launching surface from both the
simple relation (\ref{eqn:relation1}), $\Omega_{0,A}$, and
the refined relation (\ref{eqn:final2}), $\Omega_{0,B}$.
Comparing the predicted values with the exact $\Omega_0$, we find
that the simple relation (\ref{eqn:relation1}) works well for the
light wind, but under predicts the rotation rate by a factor of two
for the more extreme, heavy wind. The large discrepancy in the latter
case comes from the neglect of the fluid energy and angular momentum
at the base of the wind in deriving relation (\ref{eqn:relation1}),
which are comparable to their magnetic counterparts for the heavy
wind. The fluid contributions are included in relation (\ref{eqn:final2}),
which yields a rotation rate accurate to within about $5\%$ in both
cases. The small errors can be further reduced if the (slow) speed
with which the wind solution is initiated and the gravitational binding
energy at the observing location are accounted for properly, which
can be done. Such refinements are not warranted, however, given the
large uncertainties in real observational data, to which we now turn.

\subsection{The Case of DG Tau}

Observational data needed to locate the wind-launching region on the
disk are available for DG Tau, thanks to the spatially resolved
spectra of several forbidden lines obtained by \citet{bac00}
using HST/STIS. The STIS slit was placed along the axis of the flow
and at three pairs of locations at distances $10$, $20$, and $30\AU$ from
the axis (one on each side). Two distinct velocity components are
detected, with the high-velocity component (HVC) concentrating near
the axis and the low-velocity component (LVC) being more laterally
extended. \citet{bac02} analyzed the extended LVC in detail,
and derived the radial (line-of-sight) velocity $v_{r,\infty}$ at
four positions along
each slit, labeled I through IV in their Fig.~7. They also found
tentative evidence for rotation from the difference in radial velocity
between the regions displaced symmetrically with respect to the axis.
The axis is inclined at $\theta \approx 38\degr$ to our
line of sight \citep{em98}, which allows us to estimate the poloidal
speed $v_{p,\infty}=v_{r,\infty}/\cos\theta$, assuming a flow
predominantly parallel to the axis. This assumption is expected to
break down badly in region I, which is the closest to the disk; it
should be accurate to $\sim 50\%$ or better in the other regions.
The estimated values of $v_{p,\infty}$ in regions II through IV are
listed in Table~\ref{tab:dgtau},
along with the deprojected toroidal velocities
computed using the numbers given in Table~1 of \citet{bac02}, which
are uncertain. The angular speed $\Omega_{0}$ in Table~\ref{tab:dgtau}
is derived from equation (\ref{eqn:final2}), and the wind-launching
radius $\varpi_{0}$ is computed for a stellar mass
$\stellarmass=0.67\solarmass$
\citep{h95}.

It is clear from Table~\ref{tab:dgtau}
that the LVC of the DG Tau wind comes from a range
in disk radius, from $\sim 0.3$ to $\sim 4\AU$. The spread in launching
radius can be understood qualitatively as follows: in each of the three
regions, the flow located closest to the axis (at the $10\AU$ distance)
has the highest poloidal speed (and thus the highest specific energy)
and the lowest, or close to the lowest, toroidal speed (and thus the
lowest specific angular momentum given its small distance). To extract
the highest energy against the weakest torque on the disk (corresponding
to the lowest angular momentum), the field line passing through the
innermost location in each region must be driven at the fastest rate
at the foot point, i.e., must be anchored closest to the central star.
Detailed calculations confirm this expectation, as shown in Fig.~{fig:dgtau},
where straight lines connecting the observing locations and their
originating points on the disk are plotted. These lines give a crude
indication of the actual streamlines. They are well behaved, with two
exceptions: the lines associated with the two outer locations in region
IV cross other lines, which is problematic. However, region IV appears
to be perturbed by a limb-brightened bubble-like structure \citep{bac00},
which may have introduced additional asymmetry to the
flow with respect to the axis, and compromised the estimates of toroidal
speed and thus the foot point locations.

We reiterate that our values of $v_{p,\infty}$ are estimated from the
radial velocities at the peak of forbidden lines, which sample only
the component of poloidal velocity parallel to the axis, $v_{z,\infty}$,
in the likely case of peak emission originating from the tangent point
of our line of sight with the wind. Since the $v_\varpi$ component is
unmeasured, our values of $v_{p,\infty}$ are actually underestimates.
This effect will be most pronounced for the least collimated part of
the flow, where $v_{p,\infty}$ can be underestimated by as much as
$\sim 50\%$. Taking this effect into account could move the outermost
launching point inward by a factor up to two.

With the rotation rate $\Omega_0$ determined, we can estimate the
Alfv\'{e}n radius $\varpi_A$ by equating the fluid angular momentum at
large distances, $\varpi_\infty v_{\phi, \infty}$, with the total
angular momentum, $L\equiv \Omega_0 \varpi_{A}^{2}$ (e.g.\ \citealt{s96}).
The estimates for $\varpi_A$ are listed in Table~\ref{tab:dgtau}
and the corresponding
``Alfv{\'e}n points'' are plotted in the right panel of Fig.~{fig:dgtau}. The
Alfv\'{e}n radius turns out to be $\sim 1.8$ to $\sim 2.6$ times the
foot point radius (except for the ``streamline'' passing through the
bubble-affected outermost location in region IV; see Table~\ref{tab:dgtau}).
The typical ratio of $\varpi_A/\varpi_0\sim 2$ implies that roughly $1/4$
of the material accreted through the disk is ejected in the extended
LVC of the DG Tau wind as per the relationship $\dot{M}_w/\dot{M}_d
\approx (\varpi_0/\varpi_A)^2$ \citep{pp92}.

\section{CONCLUSIONS AND DISCUSSION}
\label{sec:conclusions}

We have developed a method of locating the launching region
of T Tauri winds, if such winds are driven magnetocentrifugally, as is
widely believed. Our method relies on the fact that the energy and
angular momentum in the wind are extracted mostly by magnetic fields
from the rotating disk, and they are related by the rate of disk
rotation because the energy extracted is the work done by the
rotating disk against the magnetic torque responsible for the
angular momentum extraction. Since most of the wind energy and angular
momentum at large, observable distances are in the measurable kinetic
form, they can be used to infer the disk rotation rate in the wind launching
region. Applying this method to the wind of DG Tau, we find that its
spatially resolved LVC comes from a region on the disk extending from
$\sim 0.3$ and $\sim 4\AU$ from the central star. This range brackets
the rough estimate of $\sim 1.8\AU$ by \citet{bac02}. It
strengthens the notion that the LVCs of T Tauri winds are driven
from relatively large disk radii \citep{kt88}.

The DG Tau wind has a HVC with a radial velocity of
$\sim 220\kms$ \citep{pyo03}.
Where this component is launched is less certain. It is unresolved
in the transverse direction in the HST observations \citep{bac00},
implying a half-width $\lesssim 5\AU$. Unless rotating at a speed much
greater than that inferred for the LVC ($\sim 10\kms$), the
HVC must be driven magnetocentrifugally from a disk region with radius
on the order of $0.1\AU$ or smaller, according to
equation~(\ref{eqn:final4}). The small launching radius is indicative
of an X-wind origin \citep{shu00}. To draw a more quantitative
conclusion on the launching radius of the HVC, its rotational speed
must be measured, which may become possible with the advent of optical
interferometry using large, ground-based telescopes.

\acknowledgments
Support for this work was provided in part by NASA grants NAG 5-7007, 
5-9180, 5-12102 and NSF grant AST 00-93091.

\begin{deluxetable}{ l c c c c c c c }
	\tabletypesize{\scriptsize}
	\tablewidth{360pt.}
	\tablecaption{Predicted $\Omega_{0}$ for Model Calculation}
	\tablehead{\colhead{Model} & \colhead{$\varpi_{0}$}
	& \colhead{$\Omega_{0}$} & \colhead{$\varpi_{\infty}$}
& \colhead{$v_{\phi,\infty}$} & \colhead{$v_{p,\infty}$}
& \colhead{$\Omega_{0,A}$} & \colhead{$\Omega_{0,B}$} \\
 \colhead{} & \colhead{($\AUns$)}
	& \colhead{(s$^{-1}$)} & \colhead{($\AUns$)}
& \colhead{($\kmsns$)} & \colhead{($\kmsns$)}
& \colhead{($\secondns^{-1}$)} & \colhead{($\secondns^{-1}$)}}
	\startdata
	L  & 0.21 & 2.16$\times 10^{-6}$ & 20.0 & 23.0 & 526.8 & 2.02$\times 10^{-6}$ & 2.11$\times 10^{-6}$ \\
	   & 0.31 & 1.23$\times 10^{-6}$ & 37.2 & 15.0 & 437.3 & 1.15$\times 10^{-6}$ & 1.20$\times 10^{-6}$ \\
	   & 0.40 & 8.18$\times 10^{-7}$ & 57.1 & 10.9 & 377.5 & 7.67$\times 10^{-7}$ & 8.02$\times 10^{-7}$ \\
	   & 0.51 & 5.66$\times 10^{-7}$ & 90.4 & \phn7.6 & 326.8 & 5.22$\times 10^{-7}$ & 5.45$\times 10^{-7}$ \\
	   \tableline
	H  & 0.21 & 2.10$\times 10^{-6}$ & 11.5 & \phn3.6 & 116.1 & 1.09$\times 10^{-6}$ & 2.13$\times 10^{-6}$ \\
	   & 0.31 & 1.21$\times 10^{-6}$ & 25.7 & \phn2.0 & \phn97.7 & 6.31$\times 10^{-7}$ & 1.19$\times 10^{-6}$ \\
	   & 0.40 & 8.06$\times 10^{-7}$ & 42.9 & \phn1.3 & \phn83.9 & 4.14$\times 10^{-7}$ & 8.38$\times 10^{-7}$ \\
	   & 0.51 & 5.60$\times 10^{-7}$ & 69.8 & \phn0.9 & \phn71.8 & 2.77$\times 10^{-7}$ & 5.55$\times 10^{-7}$ \\
	\enddata
	\label{tab:model}
\end{deluxetable}

\begin{deluxetable}{ l c c c c c c c }
	\tabletypesize{\scriptsize}
	\tablewidth{325pt.}
	\tablecaption{Predicted $\Omega_{0}$, $\varpi_{0}$ and $\varpi_{A}$
	for LVC of DG Tau Wind}
	\tablehead{\colhead{Region} & \colhead{$\varpi_{\infty}$} &
\colhead{$v_{\phi,\infty}$} & \colhead{$v_{p,\infty}$} &
\colhead{$\Omega_{0}$} & \colhead{$\varpi_{0}$} & \colhead{$\varpi_A$} &
\colhead{$\varpi_A/\varpi_0$} \\
\colhead{} & \colhead{($\AUns$)} &
\colhead{($\kmsns$)} & \colhead{($\kmsns$)} &
\colhead{($\secondns^{-1}$)} & \colhead{($\AUns$)} & \colhead{($\AUns$)} &
\colhead{}}
	\startdata
	II & 10 & \phn7.3 & 58.2 & 2.7$\times 10^{-7}$ & 0.71 & \phn1.34 & 1.9\\
	          & 20 & 13.2 & 39.5 & 2.9$\times 10^{-8}$ & 3.15 & \phn7.80 & 2.5\\
	          & 30 & \phn8.9 & 33.9 & 2.1$\times 10^{-8}$ & 3.91 & \phn9.22 & 2.4\\
	          \tableline
	III & 10 & \phn5.8 & 68.5 & 4.8$\times 10^{-7}$ & 0.48 & \phn0.90 & 1.9\\
	           & 20 & \phn5.4 & 56.5 & 1.5$\times 10^{-7}$ & 1.05 & \phn2.19 & 2.1\\
	           & 30 & \phn9.1 & 47.3 & 3.6$\times 10^{-8}$ & 2.71 & \phn7.12 & 2.6\\
	          \tableline
	IV & 10 & \phn4.2 & 87.2 & 1.1$\times 10^{-6}$ & 0.28 & \phn0.51 & 1.8\\
	          & 20 & \phn6.5 & 78.8 & 2.2$\times 10^{-7}$ & 0.83 & \phn1.99 & 2.4\\
	          & 30 & 15.6 & 55.2 & 2.7$\times 10^{-8}$ & 3.28 & 10.76 & 3.3\\
	\enddata
	\label{tab:dgtau}
\end{deluxetable}

\begin{figure}
        \begin{center}
        \plotone{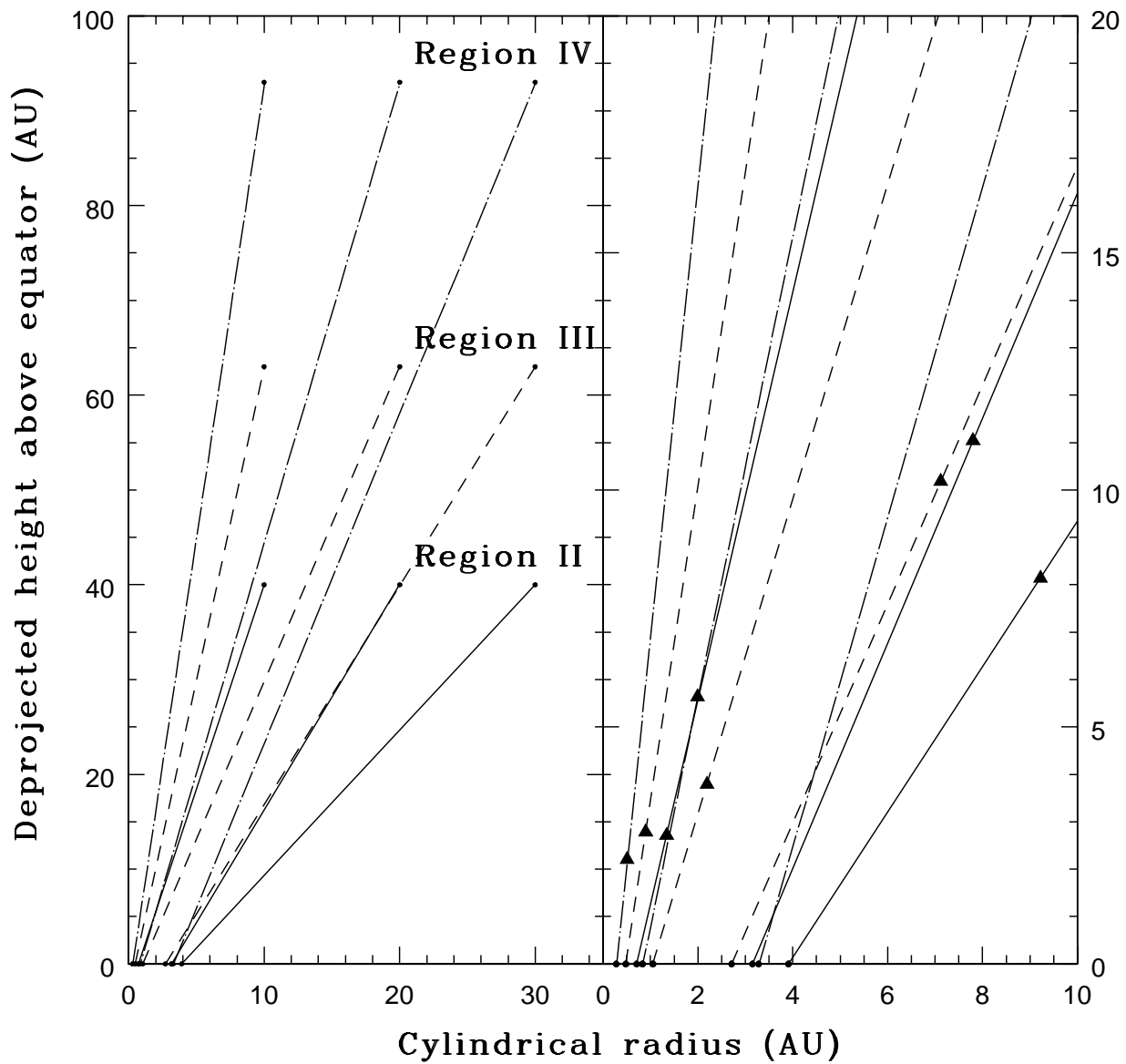}
 \end{center}
 \caption{Calculated ``streamlines'' for DG Tau.  The left panel shows
the observation points from Region II (solid lines), III (dashed), and
IV (dash-dotted) of \citet{bac02} connected to the calculated foot points
of the flow.  The right panel is a blow-up of the inner region of the flow,
showing where the flow originates from the disk.  Also shown on the right
panel is the location of the Alfv\'{e}n surface along each ``streamline''
as filled triangles.}
 \label{fig:dgtau}
\end{figure}

\end{document}